\documentclass[preprint,twocolumn]{emulateapj}

\usepackage{graphicx}
\usepackage{amsmath}
\usepackage{color}

\shorttitle{Jupiter Trojan UV spectroscopy}
\shortauthors{Wong et~al.}

\begin{document}

\title{\textit{Hubble} Ultraviolet Spectroscopy of Jupiter Trojans}
\author{Ian Wong\altaffilmark{1}\altaffilmark{$\dag$}, Michael E. Brown\altaffilmark{2}, Jordana Blacksberg\altaffilmark{3}, Bethany L. Ehlmann\altaffilmark{2,3}, and Ahmed Mahjoub\altaffilmark{3}}
\affil{\textsuperscript{1}Department of Earth, Atmospheric, and Planetary Sciences, Massachusetts Institute of Technology,
Cambridge, MA 02139, USA; iwong@mit.edu\\
	\textsuperscript{2}Division of Geological and Planetary Sciences, California Institute of Technology,
Pasadena, CA 91125, USA \\
    \textsuperscript{3}Jet Propulsion Laboratory, California Institute of Technology, Pasadena, CA 91109, USA \\
	\textsuperscript{$\dag$}\textit{51 Pegasi b} Postdoctoral Fellow
	}


\begin{abstract}
We present the first ultraviolet spectra of Jupiter Trojans. These observations were carried out using the Space Telescope Imaging Spectrograph on the \textit{Hubble Space Telescope} and cover the wavelength range 200--550~nm at low resolution. The targets include objects from both of the Trojan color subpopulations (less-red and red). We do not observe any discernible absorption features in these spectra. Comparisons of the averaged UV spectra of less-red and red targets show that the subpopulations are spectrally distinct in the UV. Less-red objects display a steep UV slope and a rollover at around 450~nm to a shallower visible slope, whereas red objects show the opposite trend. Laboratory spectra of irradiated ices with and without H$_{2}$S exhibit distinct UV absorption features; consequently, the  featureless spectra observed here suggest H$_{2}$S alone is not responsible for the observed color bimodality of Trojans, as has been previously hypothesized. We propose some possible explanations for the observed UV-visible spectra, including complex organics, space weathering of iron-bearing silicates, and masked features due to previous cometary activity. 
\end{abstract}

\section{Introduction}

The Jupiter Trojans lie at the nexus of several fundamental questions in planetary science. While the classical paradigm of solar system formation, in which the planets largely accreted \textit{in situ}, posits that the Trojans formed from the material in the protoplanetary disk immediately adjacent to the growing Jupiter, more recent models envision a drastically different scenario. These so-called dynamical instability models describe how the Trojans initially formed in the outer Solar System, beyond the primordial orbits of the ice giants, and were subsequently transported inward and captured into resonant orbits by Jupiter during a period of chaotic planet migration \citep{morbidelli,tsiganis,gomes}. Such a formation and evolution history predicts a common progenitor population for both Trojans and Kuiper Belt objects (KBOs). It follows that studying the composition of Trojans provides a direct empirical test of these dynamical instability models.

Decades of Trojan spectroscopy have been frustrated by absolutely featureless spectra in the visible and near-infrared through 2.5~$\mu$m \citep[e.g.,][]{jewitt,emerybrown,dotto,fornasier,emery}. In particular, no water ice or silicate absorption features have been detected, even though they are expected to be the predominant components in the bulk composition. Meanwhile, it has been established that Trojans have two subpopulations with different spectral properties. The bifurcation is most apparent when looking at the visible spectral slope distribution, where two peaks are evident \citep{roig,wong,wong2}. These subpopulations, which are referred to here as less-red (LR) and red (R), add an extra layer of complexity in our understanding of Trojan composition.

Fastforwarding to more recent work, a study of KL-band spectra revealed the first incontrovertible absorption feature centered around 3.1~$\mu$m on several Trojans \citep{brown}. This feature has typically been associated with fine surface water frost, but is also consistent with an N-H stretch feature attributable to ammoniated clays. The latter scenario would directly point to an outer solar system origin for Trojans.

In the absence of concrete constraints on surface chemistry from spectroscopy, many groups have proposed compositional models. A recent hypothesis \citep{wong3}, developed within the framework of dynamical instability models, describes how Trojans may have formed in the outer Solar System with roughly cometary composition. Subsequent differential sublimation of volatile ice species created distinct surface ice mixtures that would have resulted in different reddish colors upon irradiation. Our simple modeling predicted that H$_2$S ice is the crucial component, the heliocentric distance-dependent loss or retention of which created a primordial bimodality in the progenitor population of both Trojans and KBOs.

\begin{deluxetable*}{cccccc}[t!]
\tablewidth{0pc}
\tabletypesize{\scriptsize}
\tablecaption{
    Observation Details
    \label{tab:obs}
}
\tablehead{
    \multicolumn{1}{c}{Object} &
    \multicolumn{1}{c}{Color} &
    \multicolumn{1}{c}{UT Start Time}                     &
    \multicolumn{1}{c}{$V$~mag}                     &
    \multicolumn{1}{c}{G230L Exposure Time}  &
    \multicolumn{1}{c}{G430L Exposure Time} 
}
\startdata

624 Hektor & R & 2018 Jul 23 08:25:10 & 14.6 & 1853 & 180 \\
659 Nestor & LR & 2018 Aug 3 01:27:45 & 15.8 & 3593\textsuperscript{a} & 1200\\
911 Agamemnon & R & 2018 Sep 8 08:56:38 & 15.5 & 1853 & 180 \\
1143 Odysseus & R & 2018 Aug 1 19:15:17 & 15.5 & 1853&180 \\
1437 Diomedes & LR & 2018 Sep 18 20:09:10 & 15.6 & 1853& 180\\
3451 Mentor & LR & 2018 Apr 18 02:23:50 & 15.8 & 1820 & 180
\enddata
\tablenotetext{a}{The G230L observations for 659 Nestor were split across two consecutive \textit{HST} orbits.}
\end{deluxetable*}

Following up on the results of this work, we carried out ice irradiation experiments on mixtures of water ice, methanol ice, ammonia ice, with or without the addition of H$_2$S ice. The resultant spectra of the ice residues largely agree with the observed reddish, dark, and featureless optical and near-infrared spectra of Trojans, even reproducing the weak absorption at around 3.1~$\mu$m \citep{mahjoub1,mahjoub2,poston}. Notably, when looking at ultraviolet (UV) wavelengths, we found a suite of deep and distinctive absorption features. 

Motivated by these findings, we carried out an observing program using the \textit{Hubble Space Telescope} (HST) to observe six large Trojans in the UV. This paper presents the first UV reflectance spectra of Trojans and discusses the implications for our understanding of their composition and formation.

\section{Observations and Data Analysis}\label{sec:obs}

We observed six Trojans as part of a Cycle 25 HST observing program (GO 15249; PI: Ian Wong) using the Space Telescope Imaging Spectrograph (STIS). The targets are the three brightest objects from the LR and R Trojan subpopulations observable during Cycle 25 --- LR: 659 Nestor, 1437 Diomedes, 3451 Mentor; R: 624 Hektor, 911 Agamemnon, 1143 Odysseus. These spectra were obtained with a 52$\times$0.2$''$ slit in two grisms, G230L and G430L, to provide full low-resolution wavelength coverage from 200 to 550~nm. The exposure times were calculated so as to ensure a minimum signal-to-noise per 2-pixel resolution element of 2--3 at 250~nm and 8--10 at 375~nm for the two grisms, respectively. The resolution elements for the two spectral settings have widths of 0.32 and 0.54~$\mu$m, respectively.

For all targets except 659 Nestor, we scheduled all of the necessary exposures within one HST orbit; for 659 Nestor, the observations were split across two consecutive orbits. We scheduled automatic wavelength calibration frames at the beginning of each observing sequence to provide the most reliable wavelength solution for each target's spectra. Table~\ref{tab:obs} lists the observation details for the six Trojan targets from our program.

The G430L observations with STIS utilize a CCD detector that has accumulated significant radiation damage from long-term exposure to the space environment. The consequences of this damage include increased dark current, more hot pixels, and reduced charge transfer efficiency (CTE). For the first target observed in our program, 3451 Mentor, we had designed the observations in a ``nominal'' way, placing the object spectrum in the center of the 1024$\times$1024 pixel array and selecting for a subarray readout to reduce data overhead. For flux deposited in the middle of the array, the charge must be transferred many times before being passed to the readout of the detector, which is located at the bottom of the array. The official \textit{calstis} data reduction pipeline includes a subroutine that applies a correction to mitigate this CTE effect; however, it requires the full array output to calculate the correction, so the subroutine is not applicable to the subarray data. The resultant spectrum shows severely reduced calibrated flux levels and low signal-to-noise. For all subsequent targets, we adjusted the spectral setup to place the spectrum closer to the readout of the detector and selected for full array readout in order to apply the CTE correction. The G230L observations utilize a Cs$_{2}$Te Multi-Anode Microchannel Array (MAMA) detector, which does not suffer from CTE effects.

All of the spectroscopic observations were processed through the \textit{calstis} pipeline, which handles bias correction (CCD data only), cosmic ray rejection (CCD data only), dark subtraction, and flat-fielding of the raw 2D images to produce processed data frames. The pipeline subsequently applies the wavelength solution derived from the contemporaneous automatic wavelength calibration exposures and produces the extracted 1D spectrum in flux units.

\subsection{Results}\label{subsec:res}
\begin{figure*}[t!]
\includegraphics[width=0.5\linewidth]{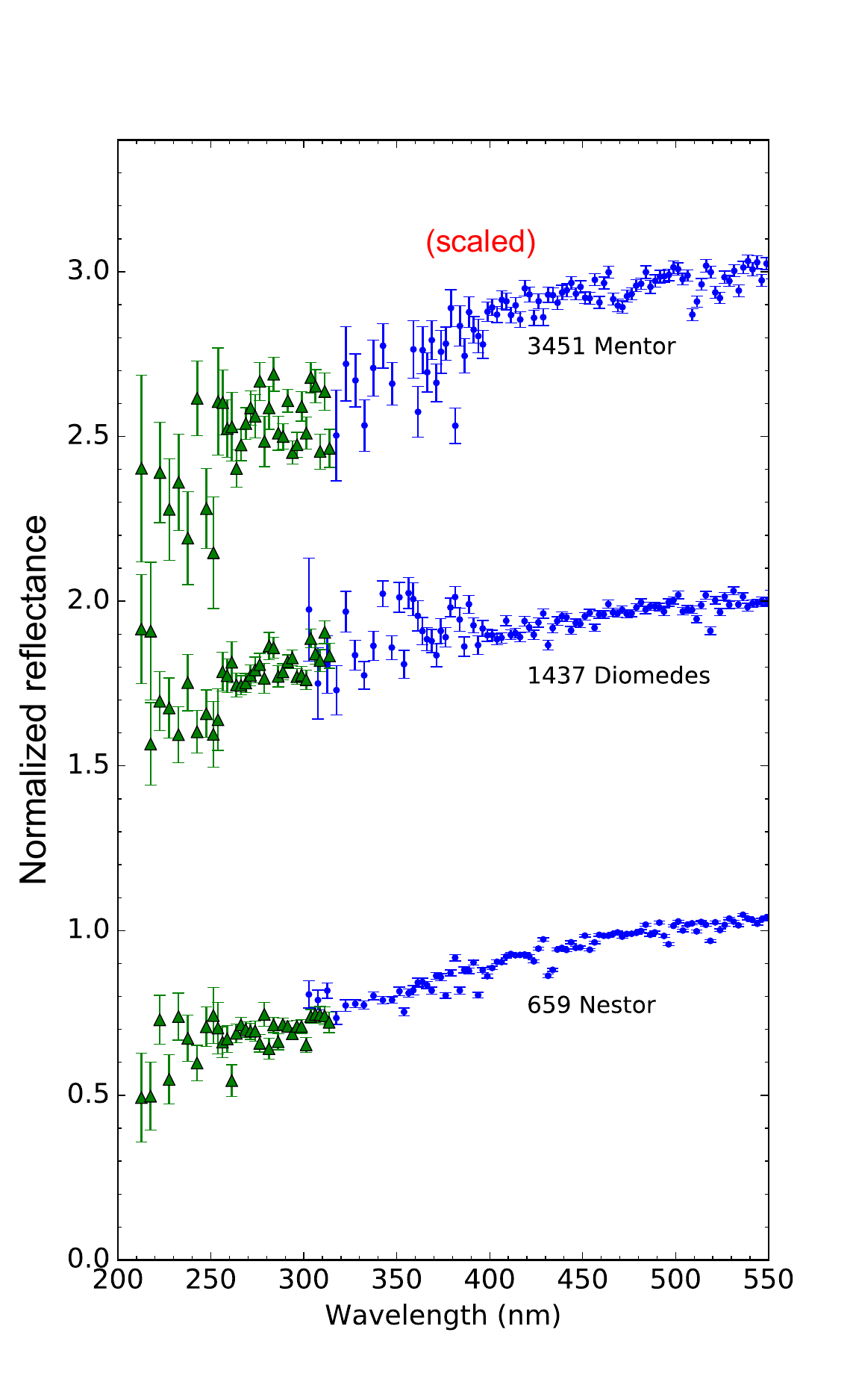}
\includegraphics[width=0.5\linewidth]{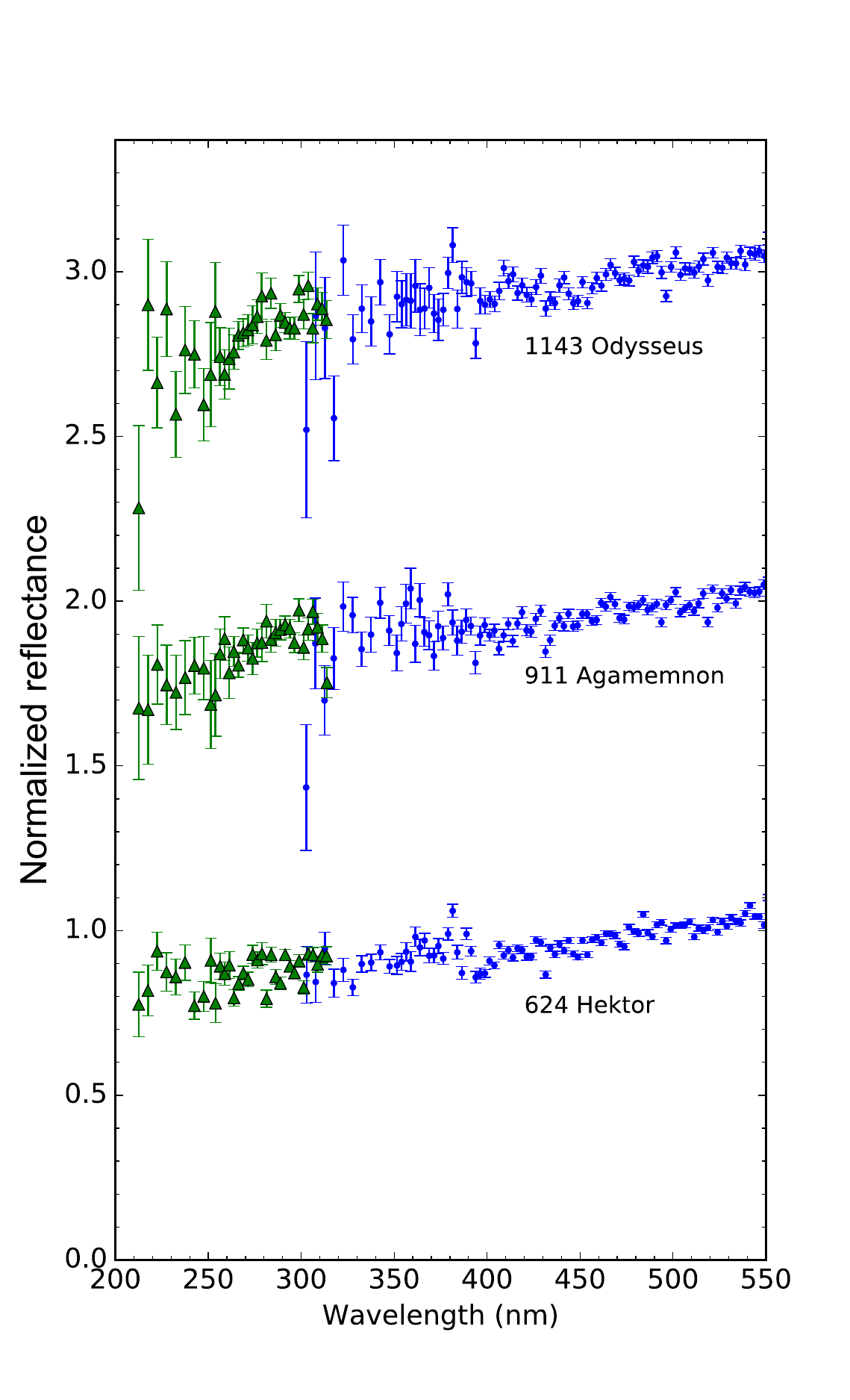}
\caption{Plot showing the binned normalized reflectance spectra of the six Trojans observed in our program, with the less-red objects on the left and red objects on the right. Each spectrum is normalized to unity at 500~nm and offset for clarity. The green triangles correspond to the spectrum taken in the G230L grism, and blue points correspond to the spectrum taken in the G430L grism. For 3451 Mentor, the G430L spectrum was affected by charge transfer efficiency losses and has been shifted up to match the trend in the G230L spectrum. None of the spectra show discernible absorption features. The error bars correspond to the binned $1\sigma$ uncertainties.}
\label{fig:spectra}
\end{figure*}

We divided the extracted 1D spectra by the ASTM E490 zero airmass solar spectrum\footnote{https://www.astm.org/Standards/E490.htm} to produce reflectance spectra. The resultant spectra for the six Trojans observed in our program are shown in Figure~\ref{fig:spectra}. All spectra have been normalized to unity at 500~nm. After applying a moving median filter with a width of 20 data points to remove $5\sigma$ outliers, we have binned the G230L and G430L spectra by 2.5~nm intervals in the ranges $\lambda\ge250$~nm and $\lambda\ge350$~nm, respectively, and by 5.0~nm intervals in the ranges $200~\mathrm{nm}<\lambda<250$~nm and $300~\mathrm{nm}<\lambda<350$~nm, respectively, where the per-pixel signal-to-noise is lower.

For all objects except 3451 Mentor, no additional normalization adjustment is needed to align the G230L and G430L spectra, since the \textit{calstis} pipeline applies an accurate flux calibration to the individual spectra, as evidenced by the consistency of the overlapping data points. In the case of 3451 Mentor, the G430L spectrum suffers from severe CTE effects, as discussed earlier. However, when comparing the shapes of the G430L spectra for the three LR Trojans, we find that the 3451 Mentor spectrum follows a similar trend as the other two. In Figure~\ref{fig:spectra}, we have shifted the G430L spectrum of 3451 Mentor up by a constant value to match the trend in the G230L spectrum. This is in essence assuming that the CTE effect is wavelength-independent; indeed, when running the CTE correction subroutine on the other spectra, the change in the spectrum is consistent with an upward shift, to first order. Since we were unable to run the subroutine in \textit{calstis} to precisely correct for the CTE effect, this shift is solely to show nominal agreement between the three LR Trojan spectra. Hereafter, we do not use the G430L spectrum of 3451 Mentor in our analysis.

The main result is that none of the individual object spectra show any incontrovertible absorption features throughout the entire wavelength range covered by our observations. Examining the shapes of the spectra, we notice a systematic difference between LR and R Trojans. While the spectra of R Trojans follow a more or less linear trend throughout the 200--550~nm region, the LR Trojan spectra show a distinct concave-down shape, with a steep UV spectral slope rolling over to a much shallower slope in the visible. The spectra of two R objects --- 911 Agamemnon and 1143 Odysseus --- display a slight break in spectral slope at around 300~nm. Fitting lines through the wavelengths ranges blueward and redward of 300~nm returns slope values that differ marginally at less than the 1.6$\sigma$ level.

\begin{figure}[t!]
\includegraphics[width=\linewidth]{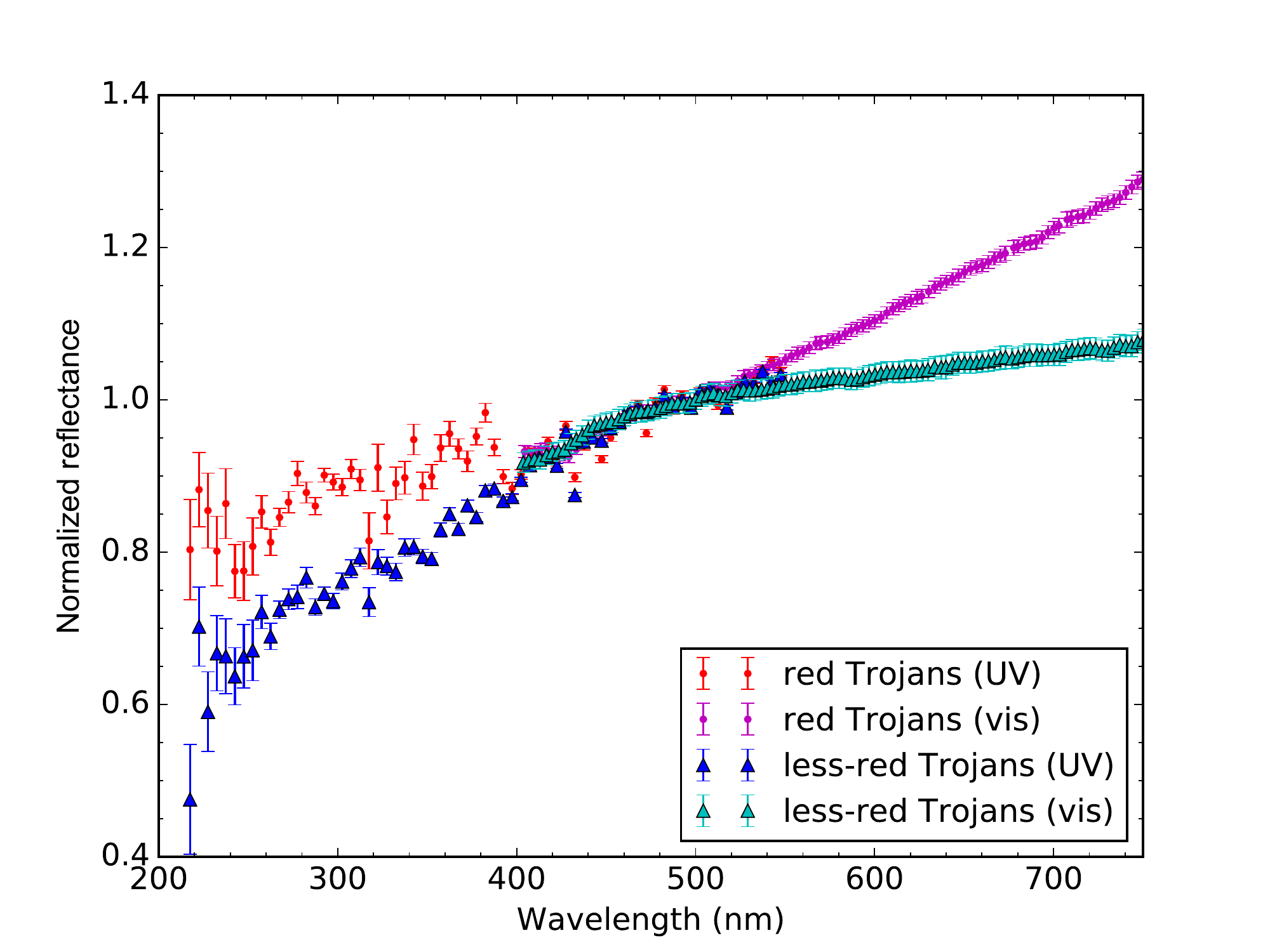}
\caption{Averaged Trojan reflectance spectra of 2 less-red and 3 red objects from our \textit{HST} STIS observations (200--550~nm), combined with published population-averaged visible spectra from \citet{emery}. While red Trojans have a shallow UV slope that transitions to a slightly steeper visible spectral slope, less-red Trojans display the opposite trend --- a steeper UV spectral slope that rolls over at $\sim$450~nm to a shallower visible slope.}
\label{fig:averages}
\end{figure}

The distinction between LR and R Trojan spectra becomes much more apparent when comparing the population-averaged spectra. In Figure~\ref{fig:averages}, we plot the average LR and R Trojan reflectance spectra, where the population-averaged visible spectra as published by \citet{emery} are also included. The HST spectra are re-binned into 5.0~nm~intervals throughout. The HST G430L spectra from our observations agree well with the trends in previously-published spectra throughout the blue visible wavelength range (400--550~nm). The difference between the LR and R Trojan spectra can be described as an anticorrelation in the spectral slopes blueward and redward of $\sim$450~nm: the R Trojan spectrum has a shallow UV slope that steepens redward of $\sim$450~nm; in contrast, the LR Trojan spectrum is significantly steeper than the R Trojan spectrum in the UV, but rolls over to a shallower linear slope in the visible.

Like the individual object spectra, the population-averaged LR and R spectra show no incontrovertible spectral features, aside from the slope change around 450~nm. On closer inspection, the previously-noted slight slope rollover in several R Trojan spectra is manifested in the population-averaged spectrum as a possible weak absorption feature centered around 240~nm. This feature is not discernible in any of the individual R Trojan spectra, with the possible exception of 1143 Odysseus. Indeed, when altering the binning widths and binning centers, the saliency of this feature varies, and the dip disappears altogether when the spectrum of 1143 Odysseus is removed from the population average. Due to the very marginal significance of this feature, we cannot claim to detect it in our observations; more spectra and/or higher signal-to-noise observations would be needed to confirm whether such a $\sim$240~nm absorption exists in the R Trojan spectra.

\section{Discussion}\label{sec:dis}

Attempting to derive detailed compositional models from such featureless spectra is ultimately a foolhardy endeavor, and the primary conclusion from the results of our HST program is that LR and R Trojans have distinct spectral properties in the UV, as they do everywhere else where spectra and/or broadband photometry are available. In this section, we briefly discuss the implications of these spectra in the context of the laboratory experiments and color hypothesis that motivated the observations and provide some tentative speculations as to the cause of the distinct UV spectral trends between LR and R Trojans.

\subsection{Laboratory spectra of ice mixtures}\label{subsec:lab}

In \citet{wong3}, we proposed a simple hypothesis to explain the origin of the LR and R Trojan subpopulations within the framework of recent dynamical instability models. We posited that when these volatile ice rich bodies were exposed to insolation after the dispersal of the gas disk, sublimation-driven loss of volatiles from the outermost layers led to the development of distinct surface compositions throughout this region. Our sublimation model showed that hydrogen sulfide ice (H$_{2}$S) would have been unstable to sublimation loss at these heliocentric distances. As a result, objects situated closer to the Sun experienced higher surface temperatures and became depleted of H$_{2}$S on their surfaces, while objects farther out retained H$_{2}$S. We predicted that the presence of H$_{2}$S would have contributed additional reddening to the surface relative to the case where H$_{2}$S was absent, leading to the development of a color bimodality among objects in the primordial planetesimal disk: LR objects without surface H$_{2}$S, and R objects with surface H$_{2}$S. The subsequent dynamical instability spread these bodies throughout the Solar System into the various present-day populations, which thereby inherited the color bimodality. Indeed, we find that Hildas, Trojans, and similarly-sized KBOs all show a robust visible color bimodality \citep[e.g.,][]{gilhutton,roig,emery,wong,wong2,wong4,wong5}.

To explore the detailed chemistry and spectral properties of the volatile ice mixtures described in the color hypothesis, we performed a series of ice irradiation experiments using the Icy World Simulation Laboratory at the Jet Propulsion Laboratory. Details of the experiments and results are described in several papers \citep{mahjoub1,mahjoub2,poston}. Two ice samples were created, following the compositional prescription detailed in \citet{wong3}: (a) a mixture of water (H$_{2}$O), ammonia (NH$_{3}$), and methanol (CH$_{3}$OH) ices, and (b) same as (a), but with the addition of H$_{2}$S ice. In order to simulate the irradiation and heating history of an icy surface during its residency in the primordial outer solar system planetesimal disk and subsequent migration to the Jupiter Trojan region, the samples were bombarded with high-energy (10~keV) electrons at 50~K, heated up to 120~K with continuing electron irradiation, and then irradiated further at 120~K. Spectroscopy of the resultant irradiated mantles revealed many of the spectral trends apparent in measured spectra of Trojans, e.g., featureless visible and near-infrared spectra through 2.5~$\mu$m, steeper visible spectral slope in the sample with sulfur, and an absorption feature at around 3.1~$\mu$m, comparable in shape to the observed feature in published KL-band Trojan spectra \citep{brown}.

\begin{figure}[t]
\includegraphics[width=\linewidth]{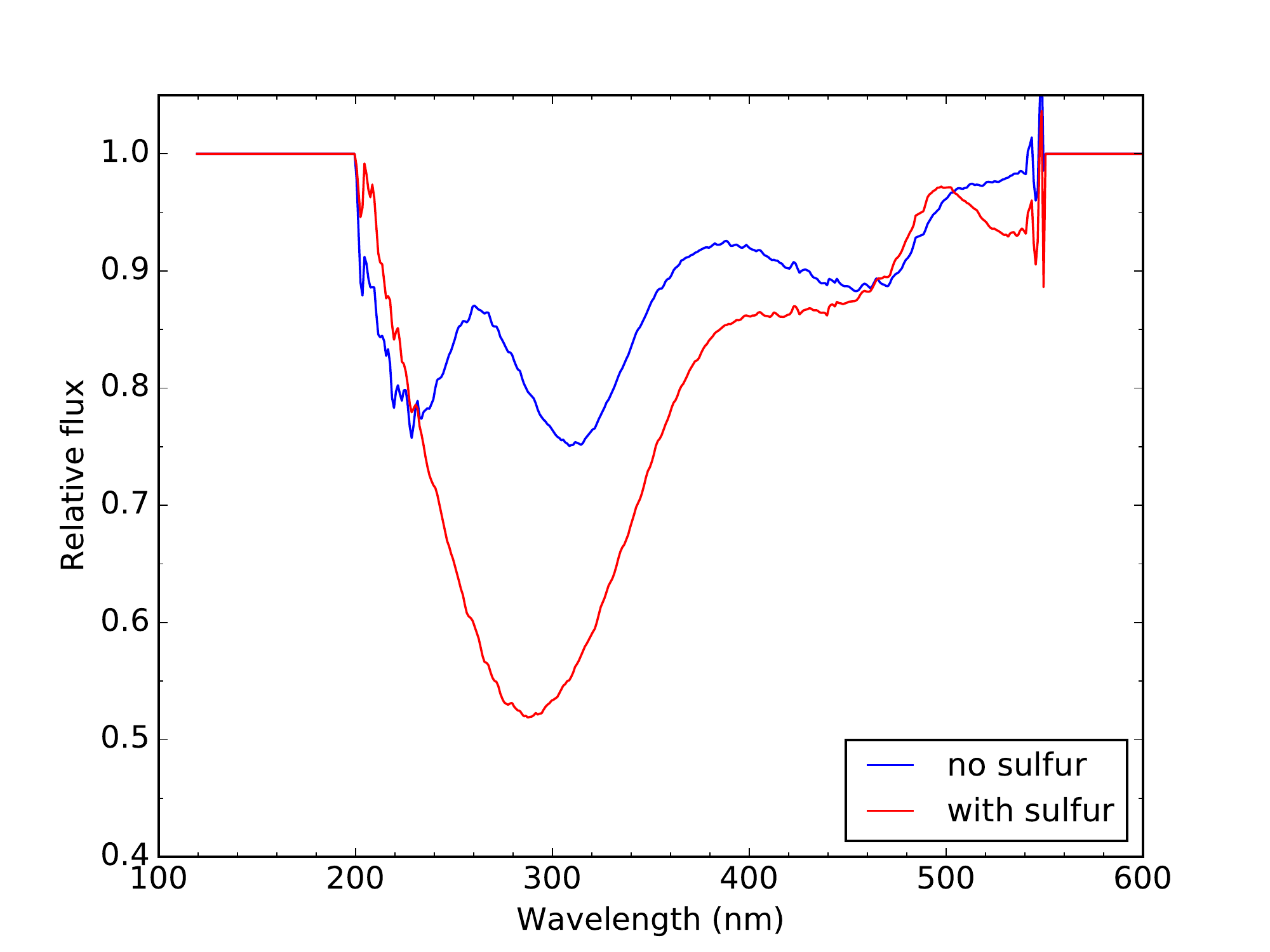}
\caption{Continuum-removed laboratory spectra of the experimental irradiated ice samples of water, methanol, and ammonia ices, with and without the inclusion H$_{2}$S ice \citep{mahjoub1,mahjoub2,poston}. The amplitude of the features across all wavelengths has been scaled so that the depth of the 3.1~$\mu$m feature in the no-sulfur sample matches the observed depth for less-red Trojans in \citet{brown}.}
\label{fig:models}
\end{figure}

In the UV, the laboratory spectra showed a number of distinguishing absorption features. In Figure~\ref{fig:models} we have plotted the continuum-removed spectra of the ice samples from our experiments after irradiation and heating. The depths of the features across all wavelengths have been approximately adjusted so that the depth of the 3.1~$\mu$m feature in the no-sulfur sample matches the observed depth for LR Trojans in \citet{brown}. The no-sulfur and with-sulfur samples, potentially corresponding to LR and R Trojans, respectively, are most discrepant between 200 and 350~nm: within this wavelength range, the with-sulfur spectrum displays a single deep absorption centered around 290~nm, while the no-sulfur spectrum shows two shallower absorption features at roughly 240 and 310~nm.

Returning to the HST UV spectra of Trojans, we see that none of the features present in the laboratory spectra are discernible in the observations to the level of noise in the data. In the context of our color hypothesis, this could mean that the ice mixtures predicted from the sublimation model are not reflective of the true surface composition of Trojans, and that the surface spectral properties of Trojans are not dominated by those of the proposed ice irradiation mantles. If this is the case, it follows that the presence or absence of sulfur irradiation chemistry is not the primary trigger for the observed color bimodality and discrepant spectral properties of LR and R Trojans.

Another possibility is that the observed UV spectra differ from the laboratory results due to a combination of grain size or relative composition effects. In our experiments, we only produced a single sample for each of the no-sulfur and with-sulfur cases, without examining the effects of varying the relative ice fractions within the mixtures or considering different grain sizes. While lowering the mass fractions of the volatile ice species (ammonia, methanol, and hydrogen sulfide) relative to water ice is expected to reduce the depth of the resultant absorption features and dilute the discrepancy between the no-sulfur and with-sulfur samples, the effects of modulating grain size can be complex and highly wavelength-dependent and as such are difficult to extrapolate from the given spectra. Further experimental study would be needed to explore these trends.

The spectral features of the icy irradiation mantles could also be masked as a consequence of outgassing. Observations of Centaurs, which are former KBOs that have been scattered inward onto giant planet crossing orbits, have revealed that a subset of them show cometary activity, particularly those with low perihelia \citep[e.g.,][]{jewittcentaurs}. If Trojans were sourced from the outer Solar System, as dynamical instability models predict, then they would have passed through an intermediate stage when their orbits were highly eccentric with low perihelia, similar to the orbits of present-day Centaurs, before being eventually captured into resonance by Jupiter. Having the same composition as KBOs, the inward-scattered Trojans would also become active, with some of the coma material settling back down on the surface and covering the irradiation mantle with a layer of fine-grained dust. Such a scenario is consistent with published thermal infrared emission spectra of Trojans, which show a broad 10~$\mu$m peak consistent with fine-grained silicates \citep{emeryir}.

\subsection{Comparison with other published UV spectra}\label{subsec:imp}

While most of our attention so far has been focused on volatile ices, more refractory components in the bulk composition of Trojans may also have a significant effect on the measured spectral properties. For example, spectroscopic studies of carbonaceous chondrites from meteorite samples have shown that several chondrite groups display featureless, reddish spectra at UV and visible wavelengths, with varying spectral slopes as well as characteristic slope rollovers from the UV to the visible similar to those seen in our HST Trojan spectra \citep[e.g.,][]{trigo}.

Much of the work exploring the effect of space weathering on the UV spectra of asteroids through observations, laboratory experiments, and modeling has been focused on main belt asteroids, particularly S- and C-type objects, as well as lunar soils. Many terrestrial iron-bearing materials (including pyroxenes and olivines) and spectra of young asteroids from collisional families show a prominent rollover in the reflectance trend from the UV to the visible between 400 and 500~nm. Meanwhile, uncollided S-type asteroids do not show a strong rollover, with mostly linear spectra throughout the UV and visible \citep{vilas1}. The range of UV and visible spectral shapes among silicate-rich bodies has been explained by the production of small amounts of nanophase iron (npFe$^{0}$) in the outermost layers of the surface regolith \citep[e.g.,][and references therein]{hendrix,markley,pieters}. The effects of npFe$^{0}$ on the reflectance spectrum include an overall reduction in albedo and the attenuation of the slope change at $\sim$450~nm, which manifests itself as a gradual reddening of the visible spectrum and a simultaneous bluing of the UV spectrum. The onset of discernible changes to the spectral shape due to space weathering occurs at mass fractions of npFe$^{0}$ as low as 0.0001\% \citep[e.g.,][]{vilas1,vilas2}.

These effects are consistent with the anti-correlated UV and visible spectral slopes in the LR and R Trojan spectra (Figure~\ref{fig:averages}). However, age and amount of space weathering cannot be invoked to explain the discrepancy between LR and R Trojan spectra, unlike in the case of asteroid families and lunar soils. The collisional rate in the Trojan asteroids is low, and the largest objects are expected to not have experienced major shattering collisions since their formation \cite[e.g.,][]{wong,wong2}. 

One possible explanation is that the relative contribution of iron-bearing minerals in the bulk composition is different between the two subpopulations. In this scenario, we may interpret the LR Trojans as containing a higher mass fraction of iron-bearing minerals than the R Trojans. The spectral signature of moderately-weathered silicates is manifested in the LR spectrum through the distinct slope rollover; perhaps this slope feature is not as attenuated as in the case of mature S-type main belt asteroids due to the significantly reduced solar wind flux at 5.2~AU when compared to 2--2.5~AU. Meanwhile, the same slope feature is diluted on R Trojans due to a greater admixture of some spectrally neutral component, such as amorphous carbon or complex organics \citep{emery}. A related explanation could be that the iron stoichiometries of the silicate component in LR and R Trojans are different, leading to different extents of space weathering effects on the UV-visible spectra. A more iron-rich silicate mineralogy would lead to an increased production of npFe$^{0}$ and a faster onset of the ensuing attenuation of the $\sim$450~nm rollover. Under this interpretation, R Trojans would be more iron-rich than LR Trojans.

The important implication of the aforementioned arguments is distinct formation environments for LR and R Trojans. Such a scenario runs counter to the predictions of recent dynamical instability models, which invariably show that all Trojans were emplaced from a single primordial reservoir of planetesimals in the outer Solar System \citep{morbidelli,roig2}. If the explanations involving different silicate fractions and/or iron stoichiometries are correct, then LR and R Trojans must have formed in two different regions of the protoplanetary disk, requiring a wholesale reevaluation of our understanding of solar system evolution.

Some organic materials also have UV-visible spectral properties that are comparable to the observed Trojan spectra. Laboratory irradiation experiments on natural and synthetic complex hydrocarbons have revealed dark, reddish visible spectra and slope rollovers in the UV, as well as distinct anti-correlated UV-visible slope changes upon irradiation. \citet{moroz} demonstrated that irradiation of asphaltine steepens the UV slope while simultaneously flattening the visible slope, producing a slope rollover between 0.4 and 0.5 microns. Irradiation experiments on polystyrene-coated olivine showed the opposite trend, with an initial steep UV and flat visible spectral shape changing to a more linear spectrum with increasing irradiation dosage \citep{kanuchova}. While previous spectral modeling of Trojans has typically limited the mass fraction of complex organics due to the absence of associated major absorption features in the near-infrared \citep[e.g.,][]{jewitt,emerybrown,dotto,fornasier,emery}, the KL-band spectra of LR Trojans show minor absorptions at 3.3 and 3.4~$\mu$m that are consistent with aromatic and aliphatic hydrocarbons, respectively \citep{brown}. 

Such organic materials can be primordial, arising within the solid material in the protoplanetary disk, or can be formed through secondary irradiation processes on the surface of small bodies, as in the case of tholins. Therefore, it is possible that differential sublimation of ice species, as described in the color hypothesis in \citet{wong3}, may still have relevance in the interpretation of Trojan surface spectral properties and the formation of the color bimodality. Ultimately, the surface chemistry of Trojans likely involves complex interactions between the icy volatile and refractory components under the action of space weathering processes that have not yet been adequately addressed in laboratory studies or spectral modeling.

\begin{figure*}
\includegraphics[width=\linewidth]{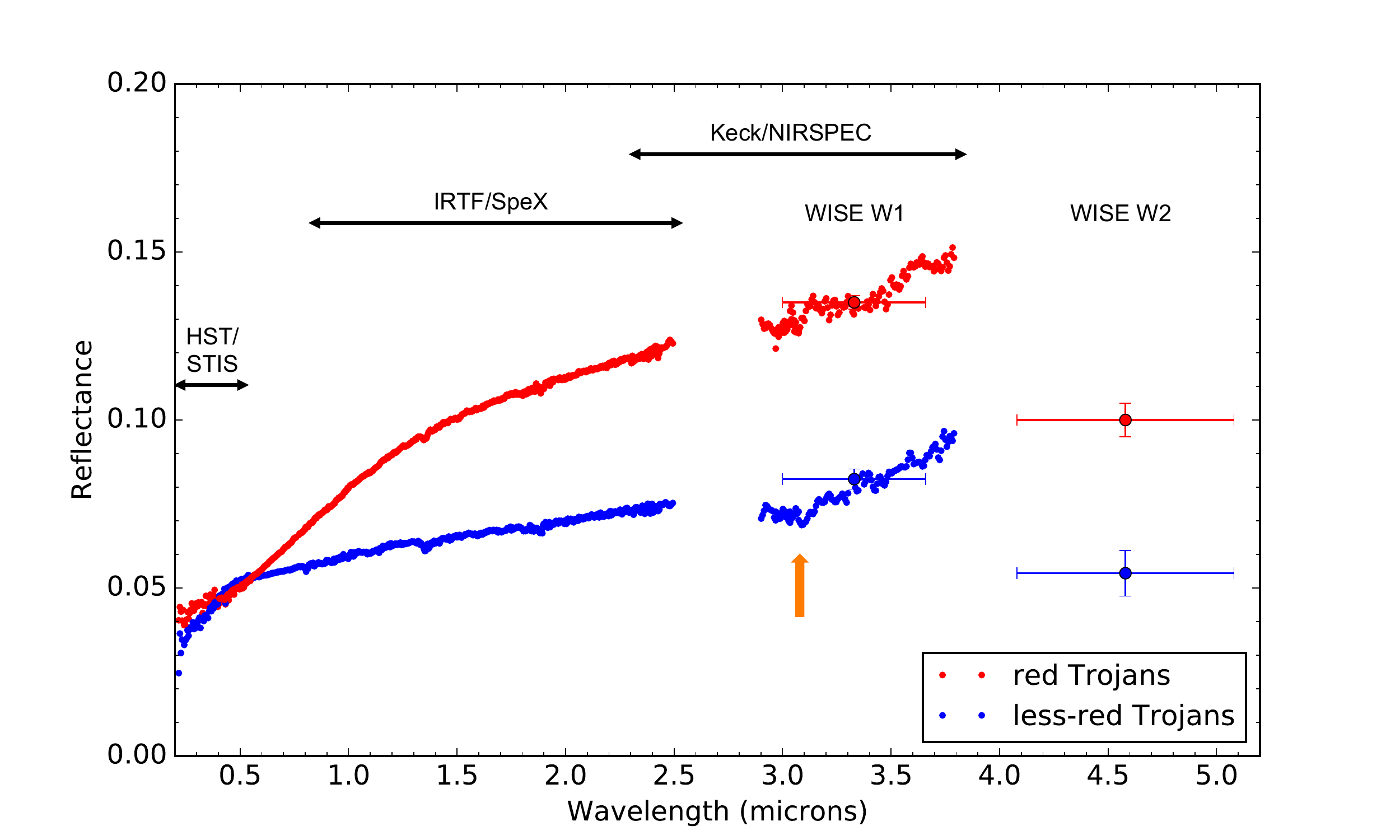}
\caption{Combined absolute reflectance plot showing the full extent of published population-averaged Trojan spectroscopy. The \textit{HST} STIS observations (200--550~nm) analyzed in this work are combined with visible spectra compiled by \citet{emery}, 0.7--2.5~$\mu$m near-infrared spectra obtained using the SpeX instrument at the Infrared Telescope Facility \citep[IRTF;][]{emery}, and K- and L-band spectra from \citet{brown}. We have also included broadband reflectance measurements obtained by the Wide-field Infrared Survey Explorer (WISE) in the W1 and W2 bands \citep[3.4 and 4.6~$\mu$m, respectively;][]{grav}. The two subpopulations are spectrally distinct throughout the full wavelength range covered by previously-published reflectance spectra. The 3.1~$\mu$m absorption in the less-red Trojan spectrum is indicated by the orange arrow. The broadband WISE reflectances suggest a deep absorption in both subpopulations between 4 and 5~$\mu$m.}
\label{fig:full}
\end{figure*}

\section{Conclusion}

In this paper, we have presented the first ultraviolet spectra of Jupiter Trojans, obtained using the STIS instrument on the \textit{Hubble Space Telescope}. The spectra of the six targets, three from each of the two Trojan color subpopulations, show no discernible absorption features throughout the wavelength range covered by our observations (200--550~nm). Examining the average spectra of less-red and red objects, we have demonstrated that the two subpopulations differ systematically in their spectral properties across the UV-visible. Specifically, red Trojans have a shallow UV spectral slope that transitions to a slightly steeper visible slope, while less-red Trojans have a steep spectral slope in the UV that rolls over at around 450~nm to a shallower visible slope. 

These ultraviolet spectra help to complete the picture of Trojan reflectance spectroscopy in wavelength ranges that are accessible from the ground or by current space-based facilities. The full absolute reflectance spectra of less-red and red Trojans are shown in Figure~\ref{fig:full}. The overall conclusion is that less-red and red Trojans have discrepant spectral properties in the ultraviolet, as they do in every other wavelength range they have been observed in. While we previously proposed a simple compositional model describing the formation of Trojans and the development of the color bimodality within the framework of recent dynamical instability models \citep{wong3}, laboratory spectra of analog ice mixtures show a suite of absorption features in the UV that are not present in the measured spectra \citep{mahjoub1,mahjoub2,poston}. Meanwhile, comparisons of the observations with other published ultraviolet spectra of minor bodies reveal similarities with carbonaceous chondrites, complex organics, as well as iron-bearing silicate materials modulated by various levels of space weathering. All in all, the composition of Trojan asteroids and the cause of the color bimodality remain poorly constrained, and further observations, spectral modeling analyses, and experimental study are needed to reach a better understanding of the specific chemical components that comprise these objects and how their spectral signatures are manifested in the observed spectra.

Building on the past body of work in Trojan spectroscopy and looking to the future, a fruitful avenue of follow-up study is to revisit wavelength ranges for which spectral features have been incontrovertibly detected in previous analyses. Primary among these is the 3--5~$\mu$m region (see Figure~\ref{fig:full}), where a broad absorption centered at around 3.1~$\mu$m has been reported, as well as putative weaker absorptions at 3.3 and 3.4~$\mu$m, likely attributable to organics; additional broadband reflectance measurements from the Wide-field Infrared Survey Explorer (WISE) in the W1 and W2 bands (3.4 and 4.6~$\mu$m, respectively) indicate the presence of deep absorption features between 4 and 5~$\mu$m \citep{grav,brown}. The upcoming launch of the \textit{James Webb Space Telescope} (JWST) will for the first time enable high-precision, high-resolution spectroscopy of minor bodies throughout the full near- and mid-infrared wavelength range. JWST will also provide exquisite capabilities to study the emission spectra of Trojans in the thermal infrared, where a suite of diagnostic features have been reported \citep{emeryir}. Comparisons of these new high-quality spectra with analogous data sets for Hildas and similarly-sized KBOs will allow for detailed evaluations of similarities and differences between the various middle and outer solar system minor body populations, with wide-ranging implications for our understanding of solar system formation and evolution.

\acknowledgements
The experimental portion of this work was conducted at the Jet Propulsion Laboratory, Caltech, under a contract with the National Aeronautics and Space Administration (NASA) and at the Caltech Division of Geological and Planetary Sciences. This work was supported in part by the Keck Institute for Space Studies (KISS). I.W. is supported by a Heising-Simons Foundation \textit{51 Pegasi b} postdoctoral fellowship. We also thank Amanda Hendrix for providing helpful referee comments that improved the manuscript.

\end{document}